\documentclass{PoS}

\usepackage{graphicx}
\usepackage{amssymb}
\usepackage{amsmath}

\makeatletter
\def\simleq{\mathrel{\mathpalette\gl@align<}}
\def\simgeq{\mathrel{\mathpalette\gl@align>}}
\def\gl@align#1#2{\lower.6ex\vbox{\baselineskip\z@skip\lineskip\z@
     \ialign{$\m@th#1\hfill##\hfil$\crcr#2\crcr\sim\crcr}}}
\makeatother

\newcommand{\Pu}{p_{\uparrow}}

\newcommand{\Nu}{n_{\uparrow}}
\newcommand{\Nd}{n_{\downarrow}}

\title{Three-Nucleon Forces explored by Lattice QCD Simulations}

\ShortTitle{Three-Nucleon Forces explored by Lattice QCD Simulations}

\author{\speaker{Takumi Doi}\\%
Center for Nuclear Study, The University of Tokyo,
Tokyo 113-0033, Japan\\
        E-mail: \email{doi@ribf.riken.jp}}

\author{for HAL QCD Collaboration}

\author{\includegraphics[width=.30\textwidth]{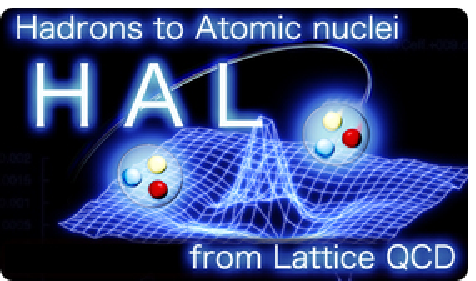}}

\abstract{
We explore three-nucleon forces (3NF)
from lattice QCD simulations.
Utilizing the Nambu-Bethe-Salpeter (NBS) wave function,
two-nucleon forces (2NF) and 3NF are determined on the same footing.
Quantum numbers of the three-nucleon (3N) system are chosen to be
$(I, J^P)=(1/2,1/2^+)$ (the triton channel).
The enormous computational cost is reduced by
employing
the simplest geometrical configuration, where
3N are aligned linearly with an equal spacing.
We perform lattice QCD simulations using 
$N_f=2$ dynamical clover fermion configurations
generated by CP-PACS Collaboration,
at the lattice spacing of $a = 0.156$ fm on a $16^3\times 32$ lattice
with a large quark mass corresponding to $m_\pi= 1.13$ GeV.
Repulsive 3NF is found at short distance.
}

\FullConference{ The XXIX International Symposium on Lattice Field Theory - Lattice 2011\\
July 10-16, 2011\\
Squaw Valley, Lake Tahoe, California}

\begin{document}

\section{Introduction}
\label{sec:intro}

Determination of the properties of three-nucleon forces (3NF)
is one of 
the most important issue
in nuclear physics and astrophysics these days.
There are various  phenomena
where 3NF
are considered to play an important role, e.g.,
the binding energies of light nuclei~\cite{Pieper:2007ax},
the properties of neutron-rich nuclei and the supernova nucleosynthesis~\cite{Otsuka:2009cs},
the nuclear equation of state (EoS)
at high density  relevant to the physics of neutron stars~\cite{Akmal:1998cf, Nishizaki:2002ih}
and
the cross sections in nucleus-nucleus elastic scatterings~\cite{Furumoto:2009zz}.
Deuteron-proton  elastic scattering experiments at
  intermediate energies have also shown a clear indication of
3NF~\cite{Sekiguchi:2011ku}.

Despite of its phenomenological importance,
microscopic understanding  of 3NF is  still limited.
Pioneered by Fujita and Miyazawa~\cite{Fujita:1957zz},
the long range part of 3NF has been commonly modeled
by the two-pion exchange (2$\pi$E),
particularly with the $\Delta$-resonance excitation.
An additional repulsive component of 3NF at short distance is 
often introduced in a purely phenomenological way~\cite{Pieper:2001ap}.
An approach based on the chiral effective field theory
is also developing~\cite{Machleidt:2011zz}.

However, since 3NF is originated by the fact that
a nucleon is not a fundamental particle,
it is most desirable to determine 3NF
directly from the fundamental degrees of freedom (DoF), i.e., quarks and gluons,
on the basis of quantum chromodynamics (QCD).
In this proceeding,
we carry out first-principle calculations of 
3NF using lattice QCD simulations.
Note that 
while 
there are lattice QCD works for 
three- and four- baryon systems%
~\cite{Yamazaki:2009ua, Beane:2009gs},
they focus on the 
energies of the multi-baryon systems,
 and extracting 3NF
is currently beyond their scope.

As for the calculation of two-nucleon forces (2NF) from lattice QCD,
 an approach based on  the 
 NBS wave function 
 has been  proposed~\cite{Ishii:2006ec, Aoki:2009ji},
so that the potential is faithful to the phase shift
by construction.
 Resultant (parity-even) 2NF 
are found to have desirable features
such as
attractive wells at long and medium
distances  and central repulsive cores at short distance.
The method has been  successfully extended to the
hyperon-nucleon (YN) and hyperon-hyperon (YY) interactions%
~\cite{Nemura:2008sp, Inoue:2010hs, Sasaki:2010bi,Inoue:2010es,Aoki:2011gt}
and meson-baryon systems~\cite{Ikeda:2011qm}.
%
%
In this report, we extend the method to the three-nucleon (3N) system,
and perform the lattice QCD simulations for 3NF 
in the triton channel,
$(I, J^P)=(1/2,1/2^+)$~\cite{Doi:2010yh, Doi:2011bw, Doi:2011gq,Doi:2011wt}.
For details of this study, refer to Ref.~\cite{Doi:2011gq}.

\section{Formalism}
\label{sec:formulation}

The detailed formulation for the determination of 2NF is 
given in Ref.~\cite{Aoki:2009ji},
and we here focus on the extension to the 3N system.
We consider the NBS wave function 
of the 3N,
$\psi_{3N}(\vec{r},\vec{\rho})$,
extracted from the six-point correlator as
\begin{eqnarray}
\label{eq:6pt_3N}
G_{3N} (\vec{r},\vec{\rho},t-t_0) 
&\equiv& 
\frac{1}{L^3}
\sum_{\vec{R}}
\langle 0 |
          (N(\vec{x}_1) N(\vec{x}_2) N (\vec{x}_3))(t) \
\overline{(N'       N'        N')}(t_0)
| 0 \rangle \\
&
\xrightarrow[t \gg t_0]{} 
&
A_{3N} \psi_{3N} (\vec{r},\vec{\rho}) e^{-E_{3N}(t-t_0)} ,
\\
\label{eq:NBS_3N}
\psi_{3N}(\vec{r},\vec{\rho}) &\equiv& 
\langle 0 | N(\vec{x}_1) N(\vec{x}_2) N(\vec{x}_3) | E_{3N}\rangle ,
\qquad
A_{3N} \equiv \langle E_{3N} | \overline{(N' N' N')} | 0 \rangle , 
\end{eqnarray}
where
$E_{3N}$ and $|E_{3N}\rangle$ denote
the energy and the state vector of the 3N ground state, respectively, 
$N$ ($N'$) the nucleon operator in the sink (source),
and
$\vec{R} \equiv ( \vec{x}_1 + \vec{x}_2 + \vec{x}_3 )/3$,
$\vec{r} \equiv \vec{x}_1 - \vec{x}_2$, 
$\vec{\rho} \equiv \vec{x}_3 - (\vec{x}_1 + \vec{x}_2)/2$
the Jacobi coordinates.

With the derivative expansion of the potentials~\cite{Murano:2011nz},
the NBS wave function can be converted to the potentials
through the following 
Schr\"odinger equation,
\begin{eqnarray}
%
\biggl[ 
- \frac{1}{2\mu_r} \nabla^2_{r} - \frac{1}{2\mu_\rho} \nabla^2_{\rho} 
+ \sum_{i<j} V_{2N} (\vec{r}_{ij})
+ V_{3NF} (\vec{r}, \vec{\rho})
\biggr] \psi_{3N}(\vec{r}, \vec{\rho})
= E_{3N} \psi_{3N}(\vec{r}, \vec{\rho}) , 
\label{eq:Sch_3N}
\end{eqnarray}
where
$V_{2N}(\vec{r}_{ij})$ with $\vec{r}_{ij} \equiv \vec{x}_i - \vec{x}_j$
denotes the 2NF between $(i,j)$-pair,
$V_{3NF}(\vec{r},\vec{\rho})$ the 3NF,
$\mu_r = m_N/2$, $\mu_\rho = 2m_N/3$ the reduced masses.
If we calculate 
$\psi_{3N}(\vec{r}, \vec{\rho})$ for all $\vec{r}$ and $\vec{\rho}$,
and if all $V_{2N}(\vec{r}_{ij})$ are obtained
by (separate) lattice calculations for genuine 2N systems,
we can extract $V_{3NF}(\vec{r},\vec{\rho})$ through Eq.~(\ref{eq:Sch_3N}).

In practice, however, the computational cost is enormous,
because of enlarged color/spinor DoF by the 3N (i.e., 9 valence quarks) 
and 
factorial number of the Wick contractions.
In order to reduce the cost, 
we first take advantage of symmetries
(such as isospin symmetry) to reduce the number of the Wick contractions.
%
Second,
we utilize that there is a freedom for the choice of a nucleon interpolating operator.
In particular, a potential is independent of the choice of a nucleon operator at the source, $N'$,
and 
we employ the non-relativistic limit operator as 
$N'=N_{nr} \equiv \epsilon_{abc}(q_a^T C \gamma_5 P_{nr} q_b) P_{nr} q_c$
with $P_{nr} = (1+\gamma_4)/2$.
Compared to the standard nucleon operator,
$N_{std} \equiv \epsilon_{abc}(q_a^T C \gamma_5 q_b) q_c$,
the spinor DoF of each (source) nucleon is reduced by half,
and the computational cost of the 3N system is reduced by 
a factor of $2^3 = 8$.
On the other hand, 
a potential is dependent on the choice of a nucleon operator at the sink, $N$,
since a NBS wave function is defined through a sink operator.
Note, however, that 
physical observables calculated from these different potentials, 
such as phase shifts and binding energies,
are unique by construction~\cite{Aoki:2009ji}.
In this sense, 
choosing $N$ corresponds to choosing the ``scheme'' 
to define the potential~\cite{Aoki:2009ji, Aoki:2010ry}.
In Ref.~\cite{Murano:2011nz}, it is found that
the non-locality of 2NF from the choice of $N=N_{std}$ is small,
and $N=N_{std}$ can be considered to be a ``good scheme''.
We therefore employ the same sink operator $N=N_{std}$
in the 3N study as well,
so that 2NF and 3NF are determined 
on the same footing.
Finally, we restrict the geometry of the 3N.
More specifically, we consider the ``linear setup''with $\vec{\rho}=\vec{0}$,
with which 3N are aligned linearly with equal spacings of 
$r_2 \equiv |\vec{r}|/2$.
%
In this setup,
the third nucleon is attached
to $(1,2)$-nucleon pair with only S-wave.
Considering the total 3N quantum numbers of 
$(I, J^P)=(1/2,1/2^+)$,
the triton channel, 
the wave function can be completely spanned by
only three bases, which can be labeled
by the quantum numbers of $(1,2)$-pair as
$^1S_0$, $^3S_1$, $^3D_1$.
Therefore, the Schr\"odinger equation
leads to 
the $3\times 3$ coupled channel equations
with the bases of 
$\psi_{^1S_0}$, $\psi_{^3S_1}$, $\psi_{^3D_1}$.
The reduction of the dimension of bases 
is expected to improve the S/N as well.
It is worth mentioning that
considering the linear setup is not an approximation:
Among various geometric components of 
the wave function of the ground state in the triton channel,
we calculate the (exact) linear setup component
as
a convenient choice to study 3NF.
While we can access only a part of 3NF from it,
we plan to extend the calculation to more general geometries
step by step,
toward the complete determination of the full 3NF.

We consider the identification of genuine 3NF.
It is a nontrivial work:
Although both of parity-even and parity-odd 2NF
are required to subtract 2NF part in Eq.~(\ref{eq:Sch_3N}),
parity-odd 2NF have not been obtained yet in lattice QCD.
(See, however, our recent progress on this issue~\cite{Murano:Lat2011}.)
In order to resolve this issue,
we consider the following channel,
\begin{eqnarray}
\psi_S \equiv
\frac{1}{\sqrt{6}}
\Big[
-   \Pu \Nu \Nd + \Pu \Nd \Nu               
                - \Nu \Nd \Pu + \Nd \Nu \Pu 
+   \Nu \Pu \Nd               - \Nd \Pu \Nu
\Big]  ,
\label{eq:psi_S}
\end{eqnarray}
which is anti-symmetric
in spin/isospin spaces 
for any 2N-pair.
Combined with the Pauli-principle,
it is automatically guaranteed that
any 2N-pair couples with even parity only.
Therefore, we can extract 3NF unambiguously 
using only parity-even 2NF.
Note that no assumption on the choice of 3D-configuration of $\vec{r}$, $\vec{\rho}$
is imposed in this argument,
and we thus can take advantage of this feature
for future 3NF calculations with various setup of 3D-geometries.

\section{Lattice QCD setup and Numerical results}
\label{sec:results}

We employ
$N_f=2$ dynamical 
configurations
with mean field improved clover fermion 
and 
RG-improved
gauge action
generated by CP-PACS Collaboration~\cite{Ali Khan:2001tx}.
We use
598 configurations at
$\beta=1.95$ and
the lattice spacing of
$a^{-1} = 1.269(14)$ GeV,
and 
the lattice size of $V = L^3 \times T = 16^3\times 32$
corresponds to
(2.5 fm)$^3$ box in physical spacial size.
For $u$, $d$ quark masses, 
we take the hopping parameter at the unitary point
as
$\kappa_{ud} = 0.13750$,
which corresponds to
$m_\pi = 1.13$ GeV, 
$m_N = 2.15$ GeV and
$m_\Delta = 2.31$ GeV.
We use the wall quark source with Coulomb gauge fixing.
In order to enhance the statistics,
we perform the measurement 
on 32 wall sources using different time slices,
and the forward and backward propagations are averaged.
The results from both of 
total angular momentum $J_z=\pm 1/2$ 
are averaged as well.
Due to the enormous computational cost,
we can perform the simulations only at a few sink time slices.
Looking for the range of sink time where the ground state saturation is achieved,
we carry out preparatory simulations for effective 2NF in the 3N system
in the triton channel at $2 \leq (t-t_0)/a \leq 11$, 
and find that the results 
are consistent with each other as long as $(t-t_0)/a \geq 7$~\cite{Doi:2011gq}.
Being on the safer side,
we perform linear setup calculations of 3NF at $(t-t_0)/a =$ 8 and 9.
We perform the simulation
at eleven physical points of the distance $r_2$. 

\begin{figure}[t]
\begin{minipage}{0.48\textwidth}
\begin{center}
%
\includegraphics[width=0.95\textwidth]{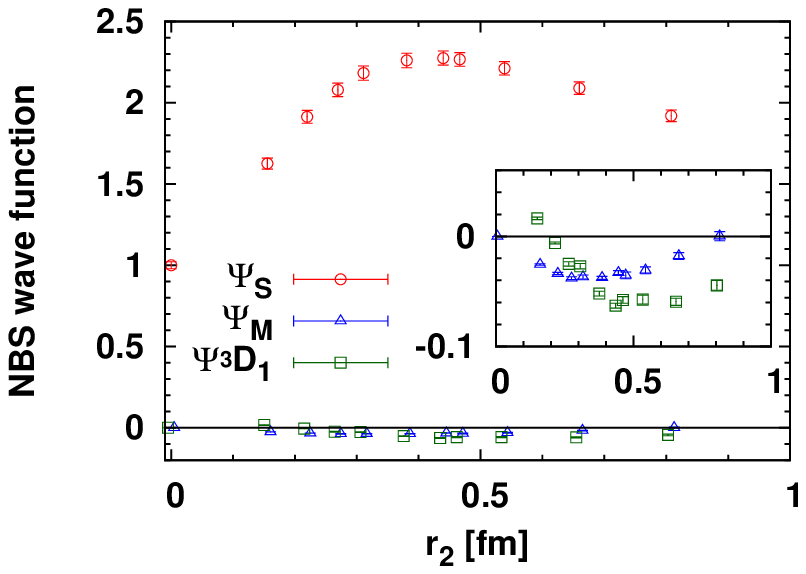}
\caption{
\label{fig:wf}
3N wave functions 
at $(t-t_0)/a=8$.
Circle (red), triangle (blue), square (green) points denote
$\psi_S$, $\psi_M$, $\psi_{\,^3\!D_1}$, respectively.
}
\end{center}
\end{minipage}
\hfill
\begin{minipage}{0.48\textwidth}
\begin{center}
%
\includegraphics[width=0.95\textwidth]{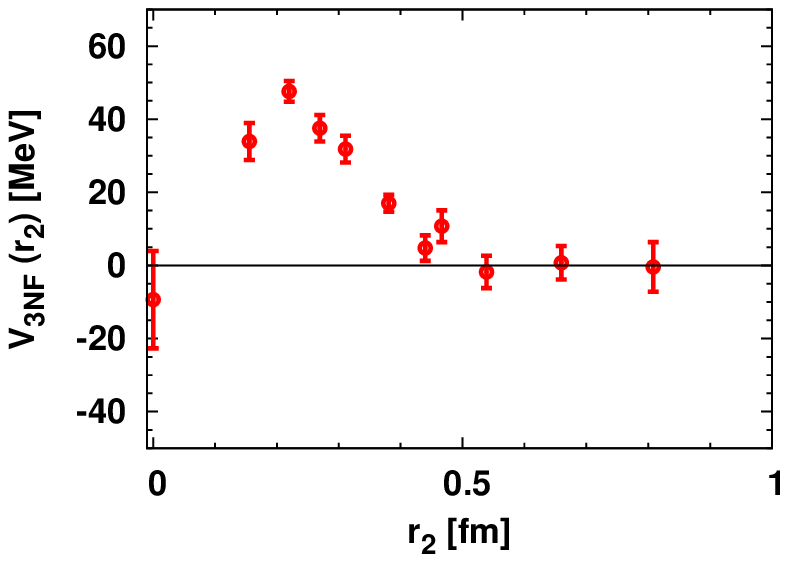}
\caption{
\label{fig:3N}
The effective scalar-isoscalar 3NF 
in the triton channel with the linear setup,
obtained at $\ \  \ (t-t_0)/a = 8$.
%
}
\end{center}
\end{minipage}
\end{figure}

In Fig.~\ref{fig:wf},
we plot
the radial part of each wave function of
$\psi_S = ( - \psi_{^1S_0} + \psi_{^3S_1} )/\sqrt{2}$,
$\psi_M \equiv ( \psi_{^1S_0} + \psi_{^3S_1} )/\sqrt{2}$
and
$\psi_{^3D_1}$ 
obtained at $(t-t_0)/a = 8$.
Here, we normalize the wave functions
by the central value of $\psi_S(r_2=0)$.
What is noteworthy is that
the wave functions are obtained with good precision,
which is quite nontrivial for the 3N system.
We observe that 
$\psi_S$ overwhelms other wave functions.
This indicates that higher partial wave components 
are strongly suppressed,
and thus the effect of the next leading order in the derivative expansion,
spin-orbit forces,
is suppressed in this lattice setup.

We determine 3NF
by subtracting 2NF from total potentials in the 3N system.
Since we have only one channel (Eq.~(\ref{eq:psi_S})) 
which is free from parity-odd 2NF,
we can determine one type of 3NF.
In this report,
3NF are
effectively represented in 
a scalar-isoscalar functional form,
which is 
often employed for the 
short-range 3NF in phenomenology~\cite{Pieper:2001ap}.

\begin{figure}[t]
\begin{minipage}{0.48\textwidth}
\begin{center}
%
\includegraphics[width=0.95\textwidth]{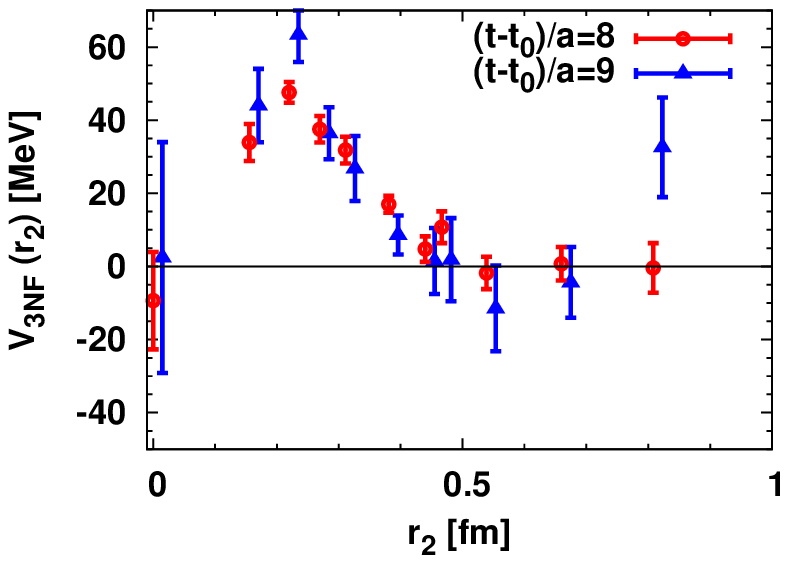}
\caption{
\label{fig:TNF_t-dep}
3NF obtained at $(t-t_0)/a = 8$ and $9$,
plotted with circle (red) points
and triangle (blue) points, respectively.
}
\end{center}
\end{minipage}
\hfill
\begin{minipage}{0.48\textwidth}
\begin{center}
%
\includegraphics[width=0.95\textwidth]{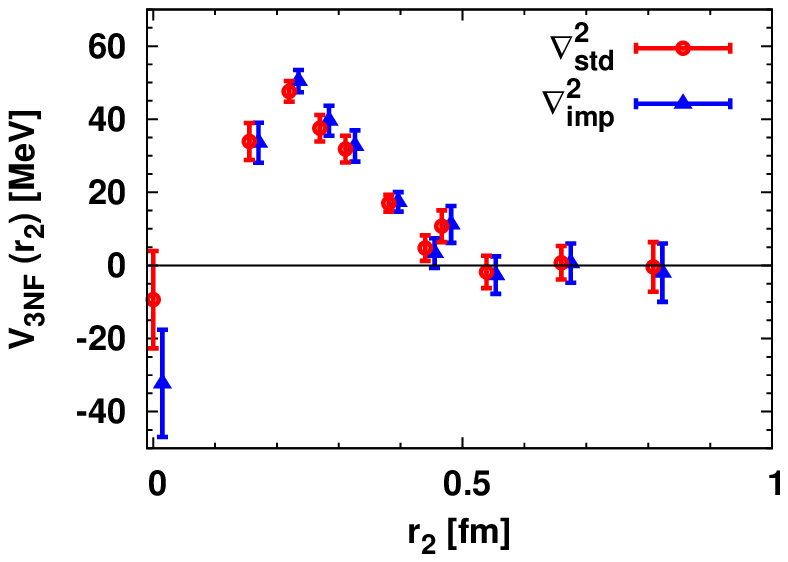}
\caption{
\label{fig:TNF_Lap}
3NF obtained at $(t-t_0)/a=8$.
Circle (red) points are obtained from $\nabla^2_{\rm std}$,
and triangle (blue) points 
from $\nabla^2_{\rm imp}$.
}
\end{center}
\end{minipage}
\end{figure}

In Fig.~\ref{fig:3N}, we plot the results
for the effective scalar-isoscalar 3NF at $(t-t_0)/a = 8$.
Here, 
we include $r_2$-independent shift by energies,
$\delta_E \simeq 5$~MeV, 
which is determined by 
long-range behavior of potentials (2NF and effective 2NF in the 3N system)~\cite{Doi:2011gq}.
While $\delta_E$ suffers from $\simleq 10$ MeV systematic error,
it does not affect the following discussions much, since $\delta_E$ merely serves as an overall offset.
In order to check the dependence on the sink time slice,
we compare 3NF from $(t-t_0)/a =$ 8 and 9 
in Fig.~\ref{fig:TNF_t-dep}.
While the results with $(t-t_0)/a=9$ suffer from quite large errors, 
they are consistent with each other within statistical fluctuations.

Fig.~\ref{fig:3N} shows that
3NF are small at the long distance region of $r_2$.
This is in accordance with the suppression
of 2$\pi$E-3NF by the heavy pion.
At the short distance region, 
however,
an indication of repulsive 3NF is observed.
Note that a repulsive short-range 3NF 
is phenomenologically required 
to explain the properties of high density matter.
Since multi-meson exchanges are strongly suppressed 
by the large quark mass, 
the origin of this short-range 3NF may be attributed to the 
quark and gluon dynamics directly.
In fact, we recall that the short-range repulsive (or attractive) cores
in the generalized two-baryon potentials 
are systematically 
calculated in lattice QCD in the flavor SU(3) limit, 
and the results are found to be well explained 
from the viewpoint of the Pauli exclusion principle in the quark level~\cite{Inoue:2010hs}.
In this context, 
it is intuitive to expect that the 3N system is subject to extra Pauli repulsion effect,
which could be an origin of the observed short-range repulsive 3NF.
Further investigation along this line is certainly an interesting subject in future.

It is in order to discuss the systematic errors in lattice simulations.
First, 
one may worry about the discretization error,
since the nontrivial results are obtained at short distance.
In particular, 
the kinetic term
(Laplacian part in the Schr\"odinger equation)
could suffer from
a substantial effect, since they are calculated by the finite difference as 
$
%
\nabla^2_{\rm std} f(x) \equiv
\frac{1}{a^2}
\sum_{i}
\left[ f(x+a_i) + f(x-a_i) - 2 f(x) \right] 
$.
In order to estimate this artifact,
we also analyze using the improved Laplacian operator
for both of 2N and 3N, 
$
\nabla^2_{\rm imp} f(x)
\equiv
\frac{1}{12a^2}
\sum_{i}
\left[
-    ( f(x+2a_i) + f(x-2a_i) )
+ 16 ( f(x+a_i)  + f(x-a_i)  )
-30   f(x))
\right] 
$.
In Fig.~\ref{fig:TNF_Lap},
we plot the comparison 
between the results of 3NF from $\nabla^2_{\rm std}$ and $\nabla^2_{\rm imp}$
at $(t-t_0)/a=8$.
It is found that 
they are consistent with each other,
and we conclude that the discretization artifact of 3NF in Laplacian operator is small.
We, however, remark that this study probes only a part of discretization errors.
Actually, the analysis with operator product expansion~\cite{Aoki:2010kx,Aoki:2010uz} shows that
2NF of $V_{2N}(r)$ tend to diverge as $r\rightarrow 0$, so 
significant discretization artifact is expected around $r=0$.
Full account of the discretization artifact can be examined by 
an explicit lattice simulation with a finer lattice,
which is currently underway.

Second, the finite volume artifact is discussed.
In this simulation, three nucleons are accommodated in (2.5 fm)$^3$ spacial lattice box.
We note that this is quite a large box
for the heavy pion, namely, $m_\pi L$ is as large as $14$.
Furthermore, the finite volume artifact 
is expected to appear mainly in large $r_2$ region,
and we avoid it as much as possible
by focusing on the short-range part of 3NF.
For the relatively large $r_2$ ($r_2 \simgeq 0.5$ fm) region,
points are carefully chosen 
so that they are located in off-axis directions.
Therefore, we expect that the finite volume artifact is not substantial
in our study.
Of course, 
a quantitative examination 
requires calculations with larger volumes,
which we defer to future studies.

Third, 
we consider the contamination from excited states.
As has been discussed, 
we do not observe sink time dependence 
for 3NF in linear setup between $(t-t_0)/a =$ 8 and 9,
nor for effective 2NF in the 3N system as long as $(t-t_0)/a \geq 7$.
It is, however, desirable to investigate 3NF in linear setup
with more sink time slices.
In particular, 
it is recently proposed~\cite{Ishii:Lat2011} 
to utilize 
the time-dependent 
Schr\"odinger equation
to further eliminate the excited state contamination.
In order to apply this method to 3NF,
linear setup calculations at additional sink time slices are in progress.

Finally, 
quark mass dependence of 3NF 
is certainly 
an important issue, 
since the lattice simulations are carried out 
only at single large quark mass.
In the case of 2NF,
short-range cores have the enhanced strength
and broaden range by decreasing the quark mass%
~\cite{Aoki:2009ji}.
We, therefore, would expect a significant quark mass dependence
exist in short-range 3NF as well.
In addition,
long-range 2$\pi$E-3NF will emerge 
at lighter quark masses, in particular, at the physical point.
Quantitative investigation through
lattice simulations with lighter quark masses
are currently underway.


We thank authors and maintainers of CPS++\cite{CPS}.
We also thank  
CP-PACS Collaboration
and ILDG/JLDG~\cite{conf:ildg/jldg} for providing gauge configurations.
The numerical simulations have been performed
on Blue Gene/L at KEK,
T2K at University of Tsukuba and SR16000 at YITP in Kyoto University.
This research is supported in part by MEXT Grant-in-Aid (20340047, 22540268),
Scientific Research on Innovative Areas (20105001, 20105003, 21105515),
Specially Promoted Research (13002001), JSPS 21$\cdot$5985 and HPCI PROGRAM,
the Large Scale Simulation Program of KEK (09-23, 09/10-24) and the collaborative
interdisciplinary program at T2K-Tsukuba (09a-11, 10a-19).


\begin{thebibliography}{99}

\bibitem{Pieper:2007ax}
  S.~C.~Pieper,
  Riv.\ Nuovo Cim.\  {\bf 31}, 709 (2008)
  [arXiv:0711.1500 [nucl-th]].

\bibitem{Otsuka:2009cs}
  T.~Otsuka, T.~Suzuki, J.~D.~Holt, A.~Schwenk and Y.~Akaishi,
  Phys.\ Rev.\ Lett.\  {\bf 105}, 032501 (2010)
  [arXiv:0908.2607 [nucl-th]].

\bibitem{Akmal:1998cf}
  A.~Akmal, V.~R.~Pandharipande and D.~G.~Ravenhall,
  Phys.\ Rev.\  {\bf C58}, 1804 (1998)
  [nucl-th/9804027].

\bibitem{Nishizaki:2002ih}
  S.~Nishizaki, T.~Takatsuka and Y.~Yamamoto,
  Prog.\ Theor.\ Phys.\  {\bf 108}, 703 (2002).\\
  T.~Takatsuka, S.~Nishizaki and R.~Tamagaki,
  Prog.\ Theor.\ Phys.\ Suppl.\  {\bf 174}, 80 (2008).

\bibitem{Furumoto:2009zz}
  T.~Furumoto, Y.~Sakuragi and Y.~Yamamoto,
  Phys.\ Rev.\  C {\bf 80}, 044614 (2009)
  [Erratum-ibid.\  C {\bf 82}, 029908 (2010)].

\bibitem{Sekiguchi:2011ku}
  K.~Sekiguchi {\it et al.},
  Phys.\ Rev.\  C {\bf 83}, 061001 (2011)
  [arXiv:1106.0180 [nucl-ex]].

\bibitem{Fujita:1957zz}
  J.~Fujita and H.~Miyazawa,
  Prog.\ Theor.\ Phys.\  {\bf 17}, 360 (1957).

\bibitem{Pieper:2001ap}
  S.~C.~Pieper, V.~R.~Pandharipande, R.~B.~Wiringa and J.~Carlson,
  Phys.\ Rev.\  C {\bf 64}, 014001 (2001)
  [arXiv:nucl-th/0102004].

\bibitem{Machleidt:2011zz}
  R.~Machleidt and D.~R.~Entem,
  Phys.\ Rept.\  {\bf 503}, 1 (2011)
  [arXiv:1105.2919 [nucl-th]]
  and references therein.

\bibitem{Yamazaki:2009ua}
  T.~Yamazaki, Y.~Kuramashi and A.~Ukawa, [PACS-CS Collab.],
  Phys.\ Rev.\  {\bf D81}, 111504 (2010)
  [arXiv:0912.1383 [hep-lat]].

\bibitem{Beane:2009gs}
  S.~R.~Beane {\it et al.},
  Phys.\ Rev.\  D {\bf 80}, 074501 (2009)
  [arXiv:0905.0466 [hep-lat]].

\bibitem{Ishii:2006ec}
  N.~Ishii, S.~Aoki and T.~Hatsuda,
  Phys.\ Rev.\ Lett.\  {\bf 99}, 022001 (2007)
  [nucl-th/0611096].

\bibitem{Aoki:2009ji}
  S.~Aoki, T.~Hatsuda and N.~Ishii,
  Prog.\ Theor.\ Phys.\  {\bf 123}, 89 (2010)
  [arXiv:0909.5585 [hep-lat]].

\bibitem{Nemura:2008sp}
  H.~Nemura, N.~Ishii, S.~Aoki and T.~Hatsuda,
  Phys.\ Lett.\  {\bf B673}, 136 (2009)
  [arXiv:0806.1094 [nucl-th]].

\bibitem{Inoue:2010hs}
  T.~Inoue {\it et al.} [HAL QCD Collab.],
  Prog.\ Theor.\ Phys.\  {\bf 124}, 591 (2010)
  [arXiv:1007.3559 [hep-lat]].

\bibitem{Sasaki:2010bi}
  K.~Sasaki [HAL QCD Collab.],
  PoS {\bf LATTICE2010}, 157 (2010)
  [arXiv:1012.5685 [hep-lat]].

\bibitem{Inoue:2010es}
  T.~Inoue {\it et al.} [HAL QCD Collab.],
  Phys.\ Rev.\ Lett.\  {\bf 106}, 162002 (2011)
  [arXiv:1012.5928 [hep-lat]].

\bibitem{Aoki:2011gt}
  S.~Aoki {\it et al.}  [HAL QCD Collab.],
  Proc. Jpn. Acad. Ser. B {\bf 87}, 509 (2011)
  [arXiv:1106.2281 [hep-lat]].

\bibitem{Ikeda:2011qm}
  Y.~Ikeda  [HAL QCD Collab.],
  in these proceedings,
  arXiv:1111.2663 [hep-lat].

\bibitem{Doi:2010yh}
  T.~Doi [HAL QCD Collab.],
  PoS {\bf LATTICE2010}, 136 (2010)
  [arXiv:1011.0657 [hep-lat]].

\bibitem{Doi:2011bw}
  T.~Doi [HAL QCD Collab.], 
  AIP Conf.\ Proc.\  {\bf 1388}, 636 (2011)
  [arXiv:1105.6247 [hep-lat]].

\bibitem{Doi:2011gq}
  T.~Doi {\it et al.} [HAL QCD Collab.],
  arXiv:1106.2276 [hep-lat].

\bibitem{Doi:2011wt}
  T.~Doi [HAL QCD Collab.], 
  Proc. of the 19th Particles and Nuclei International Conference (PANIC11),
  arXiv:1109.4748 [hep-lat].

\bibitem{Murano:2011nz}
  K.~Murano, N.~Ishii, S.~Aoki and T.~Hatsuda,
  Prog.\ Theor.\ Phys.\  {\bf 125}, 1225 (2011)
  [arXiv:1103.0619 [hep-lat]].

\bibitem{Aoki:2010ry}
  S.~Aoki,
  arXiv:1008.4427 [hep-lat].

\bibitem{Murano:Lat2011}
  K.~Murano [HAL QCD Collab.], 
  in these proceedings.

\bibitem{Ali Khan:2001tx}
  A.~Ali Khan {\it et al.}  [CP-PACS Collab.],
  Phys.\ Rev.\  D {\bf 65}, 054505 (2002)
  [E:\  D {\bf 67}, 059901 (2003)].

\bibitem{Aoki:2010kx}
  S.~Aoki, J.~Balog and P.~Weisz,
  JHEP {\bf 1005}, 008 (2010)
  [arXiv:1002.0977 [hep-lat]].

\bibitem{Aoki:2010uz}
  S.~Aoki, J.~Balog and P.~Weisz,
  JHEP {\bf 1009}, 083 (2010)
  [arXiv:1007.4117 [hep-lat]].

\bibitem{Ishii:Lat2011}
  N.~Ishii [HAL QCD Collab.], 
  in these proceedings.

\bibitem{CPS}
Columbia Physics System (CPS), 
\verb|"http://qcdoc.phys.columbia.edu/cps.html"|

\bibitem{conf:ildg/jldg}
  \verb|"http://www.lqcd.org/ildg"|,
  \verb|"http://www.jldg.org"|

\end{thebibliography}
\end{document}